\documentclass[letter,twocolumn]{jpsj2}

\def\runauthor{Yasumasa Hasegawa}
\title{
Dipole Interaction in Triplet Superconductivity with Horizontal
Line Nodes;
Orientation of the Superconducting Order Parameter in Sr$_2$RuO$_4$
}

\author{Yasumasa Hasegawa
}
\inst{
Department of Material Science,
Graduate School of Science,\\
Himeji Institute of Technology
 Ako, Hyogo 678-1297, Japan}
\recdate{June 12, 2003}
\abst{
Dipole interaction in the triplet superconductivity
is studied.
If the Cooper pairs are formed by the electrons at $\mib{r}$ and
$\mib{r}+(\pm\frac{a}{2},\pm\frac{a}{2},\pm\frac{c}{2})$ 
in the body centered tetragonal lattice, (in this case the 
line nodes run horizontally on the cylindrical Fermi 
surface),
the dipole energy  is low when the ${\mib{d}}$-vector 
is perpendicular to the direction of 
the angular momentum of the Cooper pairs ($\mib{l}$-vector).
This result is in contrast with the dipole energies in 
the ABM state of superfluid $^3$He, where $\mib{d}$-vector 
is forced to be parallel or antiparallel to 
the $\mib{l}$-vector by the dipole interaction.
The recent NQR experiment in Sr$_2$RuO$_4$ by Ishida et al. can be explained
by this result. 
}

\kword{%
triplet superconductivity, line nodes,  
Sr$_2$RuO$_4$, dipole interaction}

\begin{document}
\maketitle

The triplet superconductivity is realized in 
Sr$_2$RuO$_4$.\cite{Maeno94,Mackenzie2003}
The key experiment confirming the triplet pairing is
the observation of the temperature-independent Knight shift\cite{Ishida98},
which means that the spin susceptibility 
is not changed in the superconducting state.
It is well known that the spin susceptibility 
is the same as the normal-state value, $\chi_N$,
 if the ${\mib{d}}$-vector
of the spin-triplet pairing is perpendicular to the
external magnetic field, while it reduces as temperature becomes 
low if ${\mib{d}}$ is not perpendicular to the
external magnetic field\cite{LeggettRMP};
\begin{equation}
\chi_{ij}(T)=\chi_N
  \left( \delta_{ij}+\frac{1-Y(T)}{1+\frac{1}{4}Z_0Y(T)}d_id_j 
  \right) ,
\end{equation}
where  $Z_0$ is the Landau parameter ($Z_0 \approx -3$ in liquid $^3$He and
Sr$_2$RuO$_4$) and 
$Y(T)$ is the Yosida function for anisotropic pairing,
\begin{equation}
  Y(T)=\int\frac{d\Omega}{4\pi}\int_0^{\infty}d\epsilon_{\bf k} \frac{1}{2k_BT}
\mathrm{sech}^2 \frac{E_{\mib{k}}}{2k_BT},
\end{equation}
where
\begin{equation}
E_{\bf k}=\sqrt{\epsilon_{\bf k}^2+|\mib{\Delta}_{\mib{k}}|^2}.
\end{equation}
Therefore, if there are no other mechanism, the $\mib{d}$-vector 
is perpendicular to external magnetic field in order to
gain the magnetic energy,
\begin{equation}
 \Delta F_{\text{magn}}=- \frac{1}{2}\chi_{ij}(T)H_i H_j + 
  \frac{1}{2} \chi_N H^2.
\end{equation}

Due to the small upper critical field ($H_{c2}$) for the $z$ direction,
Knight shift in the superconducting state with $\mib{H}$ parallel to 
$c$ axis 
had not been observed.
Recently, Ishida et al.\cite{Ishida2003}
 have succeeded to observe the Knight shift 
in the superconducting state of Sr$_2$RuO$_4$ by using NQR 
with small external field along $c$ axis ($H \approx 500$G).
They observed that Knight shift does not change from the value in the normal state.
This observation may be understood by assuming that
$\mib{d}$-vector can move perpendicular to the magnetic field
even at the strength of $500$G.
In this paper we show another explanation that
the $\mib{d}$-vector is in the $x$-$y$ plain due to the dipole interaction.

In this paper we use the Leggett's notation,\cite{LeggettRMP}
 i.e.
the order parameter and the energy gap in the unitary state is given by
\begin{align}
  F_{\alpha,\beta}(\mib{r}) &\equiv 
  \langle \Psi_\alpha(\mib{R}) \Psi_\beta(\mib{R}+\mib{r}) \rangle ,\\
  \mib{F}(\mib{r})&=-\frac{i}{2}\sum_{\alpha,\beta=\uparrow,\downarrow}
    (\sigma_2 \mib{\sigma})_{\alpha\beta} F_{\alpha,\beta}(\mib{r})
 \nonumber \\
   &=\sum_{\mib{k}}\mib{F}_{\mib{k}} \exp (i\mib{k}\cdot\mib{r}),\\
 \mib{F}_{\mib{k}}&=\frac{\mib{\Delta}_{\mib{k}}}
         {2 E_{\mib{k}}} \tanh \frac{E_{\mib{k}}}{2k_B T},
\end{align}
where $\sigma_1$, $\sigma_2$, $\sigma_3$ are Pauli matrixes..

The spin-triplet paring has been  studied comprehensively in
superfluid $^3$He. In that case the Fermi surface is spherical and
the order parameter is  the $p$-wave state with very small mixing of higher-wave
(such as $f$-wave) components.
The identification of the order parameter is well established by the shift of
the NMR frequency and existence of the collective modes;
A phase is identified as the Anderson-Brinkman-Morel (ABM) state,
B phase as the Ballian-Welthermer (BW) state, and A$_1$ phase as
the non-unitary state with pairs of only one component of spins.
The ABM state breaks the time-reversal symmetry and 
it is the so-called equal-spin paring state, i.e. 
the Cooper pairs consist of the parallel  spins by taking the suitable 
direction of the quantization axis in the spin space.
In the ABM state
the  direction of the order parameter in the vector notation 
($\mib{d}$-vector) is independent of  the wave number,
\begin{equation}
  \mib{F}_{\mib{k}}^{(ABM)} = \Psi(T) \hat{\mib{d}} \sqrt{\frac{3}{2}}
  \frac{k_x + i k_y}{k} .
\end{equation}
and 
the direction of  $\hat{\mib{d}}$ is controlled by the
 external magnetic field and the dipole interaction
(and by the superfluid current, which we do not consider in this paper).

Leggett\cite{LeggettRMP} has shown that 
the dipole interaction plays important role in the triplet 
pairing.
Dipole interaction breaks a spin-orbit symmetry in the triplet superconductivity.
The dipole interaction is given by
\begin{align}
  & \hat{H}_D \nonumber \\
&= \frac{1}{2}  \sum_{\mib{r},\mib{r}'}
  \left\{
  \frac{\mib{\mu} \cdot \mib{\mu}'}
        {|\mib{r}-\mib{r}'|^3}
 -\frac{3 \mib{\mu}  \cdot (\mib{r}-\mib{r}')
          \mib{\mu}' \cdot (\mib{r}-\mib{r}')}
       { |\mib{r}-\mib{r}'|^5} 
 \right\},
\end{align}
where $\mib{\mu}$ and  $\mib{\mu}'$  are
 the operator of the magnetic moment at $\mib{r}$ and $\mib{r}'$, respectively.
The dipole energy per unit volume from  triplet  Cooper pairs is written as 
\begin{equation}
  H_D^{\text{(triplet pairs)}} = - {2} |\mib{\mu}|^2 \sum_{\mib{r}} 
\frac{|\mib{F}(\mib{r})|^2-3 |\hat{\mib{r}} \cdot \mib{F}(\mib{r})|^2}
     {r^3},
\label{dipole2}
\end{equation}
where $\hat{\mib{r}}=\mib{r}/r$.
For the pure $p$-wave pairing, the contribution of the
Cooper pairs in the dipole energy per unit volume is given by\cite{LeggettRMP}
\begin{align}
  &H_D^{\textrm{(pure $p$ wave)}}
 \nonumber \\
 & =g_D(T)
  \int \frac{d\Omega}{4\pi}
  \left\{  |\hat{\mib{d}}(\hat{\mib{k}})|^2
  - 3 |\hat{\mib{k}} \cdot \hat{\mib{d}}(\hat{\mib{k}}) |^2
    \right\},
\end{align}
where $\hat{\mib{k}}=\mib{k}/k$ is the unit vector parallel to $\mib{k}$,
the integration should be done for the direction of $\hat{\mib{k}}$ and
\begin{equation}
 g_D(T)=2{\pi}|\mib{\mu}|^2 |\Psi(T)|^2.
\end{equation}
For the ABM state the dipole energy depends on the 
angle between the $\mib{d}$-vector and the angular momentum of the Cooper pairs
($\mib{l}$-vector) as 
\begin{equation}
   H_{D}^{\textrm{ABM}}
  = g_D(T) \left( 1-\frac{3}{5} \left( \hat{\mib{d}}\cdot
 \hat{\mib{l}}\right)^2 \right) .
\end{equation}
Since  $\mib{l}$  is perpendicular to the boundary surface,
$\mib{d}$  is forced to be perpendicular to the surface
by the dipole force.
Near the transition temperature (Ginzburg-Landau  region), 
 $g_D(T)$ is estimated for superfluid $^3$He as
 $g_D^{(\textrm{$^3$He})}(T)\approx 10^{-3}(1-\frac{T}{T_c}) 
  \text{erg/cm$^3$}$.
By using the estimation of orientational energy due to the anisotropy of the
spin susceptibility,
%\begin{equation}
$  \Delta F_{\textrm{magn}}^{\textrm{($^3$He ABM)}} \approx
 5\times 10^{-7} (1-\frac{T}{T_c})$ $(\hat{\mib{d}}\cdot \mib{H})^2
\textrm{ erg/(cm$^3$ G$^2$)}$,
%\end{equation}
Leggett obtained that the orientational energy due to the dipole interaction
 corresponds to the energy due to the the external magnetic field of the order of
50 G.

For Sr$_2$RuO$_4$ we use
$\mu = 9.27 \times 10^{-21}$ erg G$^{-1}$, 
$T_c=1.5$K,
$N(0)=8.4 \times 10^{34} \text{~erg}^{-1}\text{~cm}^{-3}$
and $\chi_N = 8.2 \times 10^{-6} \text{~emu} \text{~cm}^{-3}$,
and obtain
\begin{align}
 \Psi(T) &\approx \sqrt{\frac{9.3}{2}} N(0)k_BT_c 
 \log \frac{1.13 \omega_c}{k_BT_c} 
\sqrt{ 1-\frac{T}{T_c}}
 \nonumber \\
& \approx 2 \times 10^{20} \sqrt{1-\frac{T}{T_c}} \text{~cm}^{-3}.
\end{align}
Although the transition temperature of Sr$_2$RuO$_4$ is 
about $500$ times higher than the transition temperature of superfluid $^3$He,
$\Psi(T) / \sqrt{1-\frac{T}{T_c}}$ in Sr$_2$RuO$_4$ is about $1/5$ of that
in superfluid $^3$He, since the density of states is much smaller in Sr$_2$RuO$_4$
 than in superfluid $^3$He.
We estimate 
$g_D(T) \approx 2 \times 10 (1-\frac{T}{T_c})$ erg/cm$^3$ and
$\Delta F_{\textrm{magn}} \approx 4 \times 10^{-5} H^2 \left( 1-\frac{T}{T_c}\right)
$erg/(cm$^3$G$^2$). 
Then the 
magnetic field  of the order of $700$  G is necessary to overcome 
the dipole energy, if they are competing.
Since both the dipole energy $H_D$ and the magnetic  energy 
$\Delta F_{\text{magn}}$ are proportional to $|\mib{\mu}|^2$,
the difference of the magnitude of the dipole moments in Sr$_2$RuO$_4$ 
and superfluid $^3$He does not cause the difference of the
typical magnetic field for competing dipole energy and magnetic  energy. 

Due to the topology of the Fermi surface, the $\mib{l}$-vector is thought to be 
aligned to the $\mib{z}$ direction in Sr$_2$RuO$_4$. 
If the magnetic  energy for 
$H \parallel \hat{\mib{z}}$ competes the dipole energy as in the ABM state, 
the above estimation seems to be inconsistent with the recent observation by 
Ishida et al.,\cite{Ishida2003} 
who observed that the Knight shift does not
depend on temperature when the magnetic field ($H \approx 500$G)
is applied in the z-direction. If the above estimation is applied,
the ${\mib{d}}$-vector cannot
completely determined by
the external magnetic field  and the spin susceptibility 
(and the Knight shift) should be decreased as temperature becomes lower than $T_c$.

We show that the dipole energy makes the
$\mib{d}$-vector not parallel to  $\mib{l}$ but perpendicular to $\mib{l}$
in the case of the order parameter with horizontal line 
nodes\cite{Hasegawa2000,Hasegawa2003,Zhitomirsky2001},
\begin{align}
&\mib{F}_{\mib{k}}^{\text{(horizontal line nodes)}}=\Psi(T) \hat{\mib{d}} 
 \nonumber \\
 & \times \left(
   \sin \frac{a k_x}{2} \cos \frac{a k_y}{2} 
 + i \cos \frac{a k_x}{2} \sin \frac{a k_y}{2} \right)
% \nonumber \\ \times 
  \cos \frac{c k_z}{2} ,
\label{orderparameter}
\end{align}
which is thought to be realized 
in Sr$_2$RuO$_4$,\cite{Hasegawa2000}
where $a$ and $c$ are the lattice constants.
This order parameter is consistent with the $D_{4h}$ symmetry
of the point group and   can explain most of the experiments such as
existence of the line nodes in the 
energy gap\cite{Nishizaki2000,Ishida2000},
breaking of the time reversal symmetry\cite{Luke98},
and very small angular dependence of the
thermal conductivity\cite{Tanatar2001,Izawa2001}.
If the spin-orbit coupling were strong, $\hat{\mib{d}}$ would be 
aligned to the $\hat{\mib{z}}$ direction from the point of view in the
lattice symmetry. If the spin-orbit coupling is small and can be neglected, however,
the direction of the ${\mib{d}}$-vector should be controlled
by the external magnetic field and the dipole interaction.
In this paper we neglect the spin-orbit coupling for the 
driving force for the orientation of the $\mib{d}$-vector.
 
Since the order parameter given in Eq.~(\ref{orderparameter})
correspond to the pair at $\mib{r}$ and 
$\mib{r}+(\pm\frac{a}{2},\pm\frac{a}{2}, \pm\frac{c}{2})$,
the dipole energy can be easily calculated from Eq.~(\ref{dipole2}).
By using
\begin{equation}
  \mib{F}^{\text{(horizontal line nodes)}}(r)
  =\frac{1}{8}\Psi(T)\hat{\mib{d}} f(\mib{r})
\end{equation}
where
\begin{equation}
  f(\mib{r})=
  \begin{cases}
    (1+i)
         &\text{if $\mib{r}=( \frac{a}{2}, \frac{a}{2},\pm\frac{c}{2})$}\\
    (-1+i) 
         &\text{if $\mib{r}=(-\frac{a}{2}, \frac{a}{2},\pm\frac{c}{2})$}\\
    (-1-i)
         &\text{if $\mib{r}=(-\frac{a}{2},-\frac{a}{2},\pm\frac{c}{2})$}\\
    (1-i)
         &\text{if $\mib{r}=( \frac{a}{2},-\frac{a}{2},\pm\frac{c}{2})$}\\
     0   &\text{otherwise}
  \end{cases}
\end{equation}
we get
\begin{align}
   & H_D^{\text{(horizontal line nodes)}}
  =- \frac{1}{\pi}g_D(T) 
 \nonumber \\
 & \times   
  \frac{v_0}{4 r_0^3}
  \left(
 1-\frac{3}{r_0^2}\left[ 
  \left(\frac{a}{2}\right)^2 \sin^2\theta 
 +\left(\frac{c}{2}\right)^2 \cos^2\theta \right]\right)
\end{align}
where $\theta=\cos^{-1}(\hat{\mib{d}}\cdot \hat{\mib{z}})$,
$v_0=\frac{1}{2}a^2c$ and 
\begin{equation}
  r_0=\sqrt{2\left (\frac{a}{2}\right) ^2+\left(\frac{c}{2}\right)^2}.
\end{equation}
Since $c>a$ in Sr$_2$RuO$_4$ ($a=0.387$nm and $c=1.274$nm),\cite{Maeno94}
the dipole energy is small when the ${\mib{d}}$-vector is in the $x-y$ plane
($\theta=\frac{\pi}{2}$).
When magnetic field is applied in the $z$ direction 
the $\mib{d}$-vector does not change the direction. As a result, 
spin susceptibility is the same as that in the normal state.
When magnetic field is applied in the $x$ direction,
the  $\mib{d}$-vector is aligned to the $y$ direction. The 
spin susceptibility in this case is also the same as that in the normal state.

On the other hand if the Cooper pair is formed in the nearest sites in the 
 $x-y$ plane, we get
\begin{equation}
   \mib{F}_{\mib{k}}^{\text{(nodeless)}}
   = \Psi \hat{\mib{d}} \left(\sin ak_x + i \sin k_y \right),
\end{equation}
and
\begin{equation}
  H_D^{\text{(nodeless)}}
  =- \frac{1}{\pi} g_D(T) 
  \frac{v_0}{2 a_0^3}
  \left(
 1-\frac{3}{2}  \sin^2\theta
   \right)
\end{equation}
In this case the dipole energy is minimized when $\mib{d}$-vector is parallel
to $\hat{\mib{z}}$, as in the ABM state.
Then the magnetic field smaller than the critical value 
($\approx 700$G estimated in Sr$_2$RuO$_4$) cannot rotate the $\mib{d}$
vector and the spin susceptibility becomes small as temperature becomes low.
%%%%%%%%%%%fig_1%%%%%%%%%
\begin{figure}[tb]
\begin{center}
\includegraphics[width=0.45\textwidth]{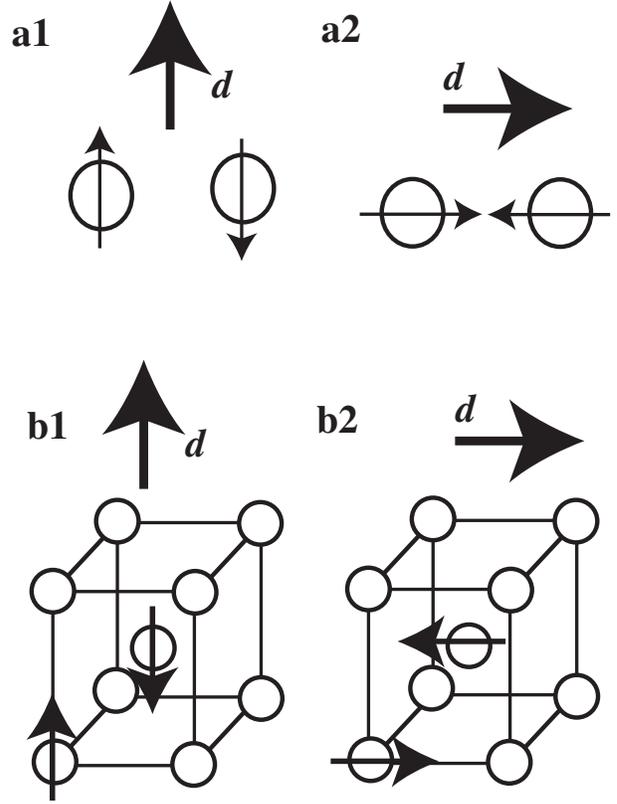}
\end{center}
 \caption{
Dipole interaction between electrons on the lattice.
Since the $d$-component of the total magnetic moment of the Cooper pair
is zero, the dipole energy for the Cooper pair in the plain perpendicular to the
$\mib{d}$-vector ($a1$) is lower then that for the
Cooper pair along the   $\mib{d}$-vector ($a2$).
For the body centered tetragonal lattice,  the Cooper
pair with $\mib{d} \parallel \hat{\mib{z}}$ ($b1$) 
has higher dipole energy than 
the Cooper pair with 
 $\mib{d} \perp \hat{\mib{z}}$ ($b2$), when $c>a$.
}
\label{fig1}
\end{figure}

The above result can be understood as follows;
along the $\mib{d}$-vector the Cooper pair is made of antiparallel spins, 
$(|\uparrow \downarrow\rangle +|\downarrow \uparrow\rangle)/\sqrt{2}$.
For the pair of spins at $\mib{r}$ and $\mib{r}+(a,0,0)$, a pair of the up and the
 down spins has lower energy than a pair of spins parallel to $+x$ and $-x$
(Fig.~\ref{fig1}, $a1$ and $a2$).
In the body-centered tetragonal lattice
  a pair of the up spin at $\mib{r}$ and the
 down spins  at $\mib{r}+(\frac{a}{2},\frac{a}{2},\frac{c}{2})$ has 
higher energy than a pair of spins $+x$ at $\mib{r}$ and $-x$ at  
$\mib{r}+(\frac{a}{2},\frac{a}{2},\frac{c}{2})$, when $c>a$ (Fig.~\ref{fig1},
$b1$ and $b2$).

The direction of the $\mib{d}$-vector can be observed by the
Josephson effect between triplet and singlet 
superconductors.\cite{Geshkenbein86,Hasegawa98}
The Josephson current between triplet and singlet superconductors
is possible due to the spin-orbit coupling in the triplet superconductivity,
if 
the conservation of the total (spin and orbital) angular momentum, 
$L_{\perp}=-S_{\perp}=\pm \hbar$, or 
the selection rule, 
$\langle \mib{n} \times \mib{k} \cdot \mib{d}(\mib{k}) \rangle \neq 0$, is satisfied,
where 
 $\mib{n}$ is the surface normal vector and
$L_{\perp}$ ($=\mib{L}\cdot \mib{n}$) and $S_{\perp}$ ($=\mib{S}\cdot \mib{n}$)
 are the 
normal components
of the total angular momentum and the total spin of the Cooper pair, respectively.
Note that if $\mib{l} \parallel \hat{\mib{z}}$, 
the average values of the $x$ and $y$ components of the
Cooper-pair's total 
angular momentum  are zero, but 
Josephson current in the $x$-$y$ plane
 is possible, since this state is the superposition of 
$L_x=\hbar$ and $L_x=-\hbar$. 
Jin et al.\cite{Jin2000} have observed the Josephson current 
between Sr$_2$RuO$_4$ and s-wave superconductor (In)  in the in-plane
direction but not along the $c$-axis.  
This result is consistent with the state that the $\mib{d}$-vector 
is along the $c$-axis.
However, we should be careful to conclude
the direction of the $\mib{d}$-vector from the Josephson junction experiments,
 because the Josephson current may be reduced by
the anisotropy of
the coherence length or the pair-breaking effect at the boundary.
The observation of 
the Josephson current along the $c$-axis is reported\cite{Sumiyama2002}, 
where the authors pointed out
the possibility of the Josephson current due to 
the existence of the steps or Ru lamellas.

%, however, are still 
%controversial.\cite{Jin99,Jin2000,Sumiyama2002}

The Fermi surface of Sr$_2$RuO$_4$ consists of three cylindrical sheets.
If the energy gap has horizontal nodes in some sheet(s) of the Fermi surface
and it has no nodes on the other sheet(s) of 
the Fermi surface\cite{Zhitomirsky2001,Yakiyama2003},
dipole force to align the $\mib{d}$-vector competes each other.
Then the net dipole force to orient the $\mib{d}$-vector may
be small.

In conclusion, the $\mib{d}$-vector is shown to be perpendicular to the 
$\mib{l}$-vector due to the dipole interaction if the order parameter
is in the form given in Eq.~(\ref{orderparameter}), which has horizontal line nodes
of the energy gap.\cite{Hasegawa2000}
Since the $\mib{l}$-vector is thought to be aligned to  the $z$ direction
in Sr$_2$RuO$_4$,  
the $\mib{d}$ vector can locate in the $x$-$y$ plane. Therefore the  magnetic
 energy does not conflict with the dipole energy, and
if the spin-orbit coupling can be neglected,
 the spin susceptibility does not depend on temperature 
regardless of the direction of
the magnetic field,
%In this paper the spin-orbit interaction is neglected.
which is consistent with
the temperature-independent Knight shift\cite{Ishida98,Ishida2003}.

The author thanks K. Ishida for showing the experimental data
 prior to publication and
valuable discussion.
He also thanks A. Sumiyama, Y. Kohori, Y. Takahashi and H. Nakano 
 for discussions. 

%\hspace{0.5cm}

%\narrowtext
%\section{Introduction}

%\vspace{-0.2cm}
%

\end{document}